\documentclass[12pt]{iopart}

\usepackage{iopams}

\usepackage{epsfig}
\def\no{\nonumber}
\def\a{\alpha}

\def\be{\begin{equation}}
\def\bea{\begin{eqnarray}}
\def\eea{\end{eqnarray}}
\def\ee{\end{equation}}
\def\bi{\begin{itemize}}
\def\ei{\end{itemize}}
\def\cross{\times}
\def\d{\delta}
\def\D{\Delta}

\def\ben{\begin{enumerate}}
\def\een{\end{enumerate}}

\begin{document}

\title[Optimal statistic for detecting inspirals with LISA...]
{Optimal statistic for
detecting gravitational wave signals from binary inspirals with LISA}

\author{Aaron Rogan and Sukanta Bose}
\address{Department of Physics and Program of Astronomy, Washington State
University, Pullman, WA 99164-2814, USA}

\ead{roganelli@wsu.edu,sukanta@wsu.edu}

\date{January 16, 2004}

\begin{abstract}
A binary compact object early in its inspiral phase will be
picked up by its nearly monochromatic gravitational
radiation by LISA. But even this innocuous appearing candidate
poses interesting detection challenges. The data that will be
scanned for such sources will be a set of three functions of
LISA's twelve data streams obtained through time-delay
interferometry, which is necessary to cancel the noise contributions
from laser-frequency fluctuations and optical-bench motions to these
data streams. We call these
three functions pseudo-detectors. The sensitivity of any
pseudo-detector to a given sky position is a function of LISA's
orbital position. Moreover, at a given point in LISA's orbit, each
pseudo-detector has a different sensitivity to the same sky
position. In this work, we obtain the optimal statistic for
detecting gravitational wave signals, such as from compact
binaries early in their inspiral stage, in LISA data. We also
present how the sensitivity of LISA, defined by this optimal statistic,
varies as a function of sky position and LISA's orbital location.
Finally, we show how a real-time search for inspiral signals can
be implemented on the LISA data by constructing a bank of templates
in the sky positions.
\end{abstract}

\maketitle
\section{\label{sec:Intro}Introduction}
The commissioning of Earth-based long-baseline gravitational wave
(GW) interferometers has finally come to fruition a little less than
three decades since the discovery of the Hulse and Taylor
binary pulsar in 1974 \cite{Hulse:1974eb} and the subsequent
confirmation of the emission of gravitational waves (GWs) by that
system.\cite{Taylor:1979a}
On the other hand, the Laser Interferometric Space Antenna (LISA) is
being designed as a space-based detector to observe low-frequency
GWs (in the milli- to deci-hertz band) to complement the
high-frequency observations (in the deca- to kilo-hertz band) of its
Earth-based counterparts. Many of the sources of low-frequency GWs
fall in the category of compact binary objects, which include white
dwarfs, neutron stars, and black holes. During the early stages of
the inspiral, these compact binaries will emit almost monochromatic
low-frequency gravitational waves, which will produce detectable
signals in the LISA data.

\begin{figure}[!hbt]
\centerline{\psfig{file=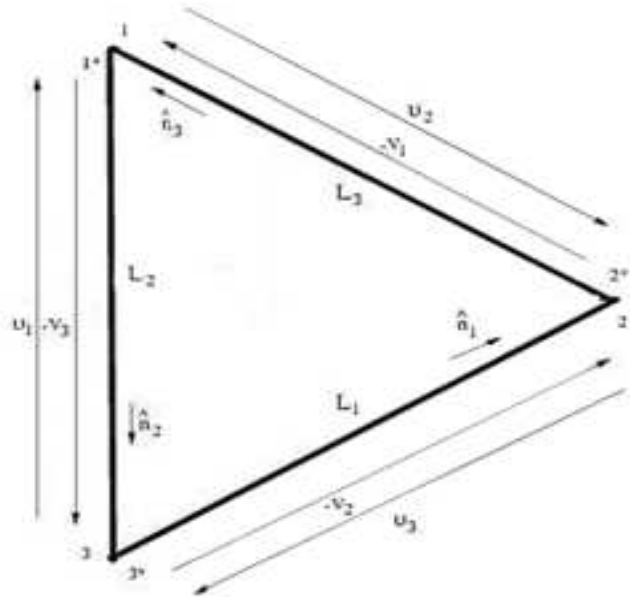,height=5.in,width=4.in}}
\caption{LISA consists of 3 spacecrafts located at the vertices of
an equilateral triangle. These craft exchange six elementary data
streams, labeled $U_i$ and $-V_i$. The  $U_i$ beams propagate
clockwise, whereas the $-V_i$ propagate counterclockwise.}
\label{LISAtri}
\end{figure}

LISA will comprise three spacecrafts located at the vertices of a
nearly equilateral triangle (as shown in Fig. \ref{LISAtri}) with
the side lengths equal to 5 million
kilometers.\cite{SysTech:2000rep} Each craft will house a couple of
laser-mounted optical benches and proof masses and will freely fall
around the Sun in an orbit that lies on a plane slightly tilted with
respect to the ecliptic. The tilt will vary from one craft to the
other. By inclining the plane of the triangle to the ecliptic by a
constant angle of 60$^\circ$, the side-length of the triangle can be
maintained at a fixed value. In the process, the triangle will
complete one spin about its normal while its centroid, trailing
behind the Earth by about 8.3 million kilometers, completes one
orbit around the Sun.

The effect of a GW on LISA will be to change the physical distances
between its freely falling proof masses. This change will be
registered as fractional frequency shifts in the six laser beams
exchanged among the three space-crafts. The shifts will have
additional contributions from various noise sources. These include
two primary ones, namely, the laser-frequency fluctuations (contributing
a fractional shift of about 10$^{-13}~{\rm Hz}^{-1/2}$) and the
optical-bench motions (with a fractional shift of about
10$^{-16}~{\rm Hz}^{-1/2}$). In order to detect the GWs (with a
strain around 10$^{-21}~{\rm Hz}^{-1/2}$), it is imperative that
these noises be mitigated by several orders of magnitude. A data
analysis technique for achieving this goal
was accomplished by Armstrong, Estabrook, and Tinto.\cite{Armstrong:99}.
They showed that by combining appropriately the time-delayed versions
of these six data streams, with six additional ones arising from laser beam
exchanges between adjacent optical benches on each craft,
one can eliminate the two primary noises, thus,
rendering the LISA data analyzable for GW signals.

In this paper, we formulate a strategy for detecting nearly monochromatic
gravitational waves from inspiraling compact binary objects in the LISA
data. There are two complementary aspects to such a strategy. The first is
to deduce the maximum number of noise-independent detectors that
LISA offers. And the second is to construct the appearance of a GW signal
in them. This allows one to match the data from these detectors with
a template of the expected signal in them.
Whether a match is strong enough
to warrant a detection is then decided based on the rate of
false alarms at the level of that match.
For the problem of detecting low-mass compact binaries, involving
white-dwarfs and neutron stars, the waveforms obtained by the quadrupole
approximation \cite{Peters:64}
suffice for accurately modeling the expected signal.
This waveform allows for a slightly ``chirping''
source, i.e., a source whose orbital frequency $\Omega_0$ is increasing
at a rate $\dot{\Omega}_0 \ll \Omega_0/T$, where $T$ is the observation
period. We will use this waveform as a template in our detection strategy.

The layout of the paper is as follows. In Sec. \ref{sec:pseudo}, we
enunciate the three noise-independent
data combinations, or ``pseudo-detectors'', that were first obtained
in Ref. \cite{Prince:2002hp} by combining the frequency shifts of the
twelve data streams exchanged among the LISA spacecrafts through time-delay
interferometry (TDI) \cite{Tinto:yr}. In Sec. \ref{sec:gwstrain}, we obtain the
form of the gravitational-wave strain caused by a compact binary source
in these three pseudo-detectors. This provides the templates required to
search for signals from such sources in the LISA data.
We then study the antenna patterns of these pseudo-detectors at
different points in LISA's orbit, emphasizing how their relative
sensitivities to a mildly chirping source vary as a function of sky
position and source frequency. The study bears out the fact that
noise-independence of detectors is a property distinct from their
geometric independence. Indeed, at GW wavelengths larger than about
0.1 AU, which we will term as the long-wavelength limit, the strain
of the third pseudo-detector tends to the difference of the strain of
the first two. We then derive the optimal statistic for detecting
gravitational waves from (non-spinning) compact binary inspirals by
coherently combining the data of the three pseudo-detectors
in Sec. \ref{sec:statistic}.

The statistic obtained here tracks the Doppler modulation of the source
frequency induced by the motion and time-varying orientation of LISA
with respect to that source.\cite{Rogan:stat}
For Gaussian noise, we derive the probability distribution of our statistic,
which can be used to compute the signal-to-noise ratio (SNR) of a candidate
event. We use this distribution to predict the behavior of false-alarm and
detection rates as a function of the detection threshold set for the
statistic.
In Sec. \ref{sec:templates}, we construct the metric on the space of
parameters that allows one to estimate the fractional loss of SNR
for a given mismatch between the template and the signal parameters.
Using this metric we esimate the number of templates that will be
required to search the full astrophysically relevant volume of the
parameter space while suffering a loss in SNR of no more than 3\%.
We also estimate the computational speed required to implement such a
search in real time.
We briefly summarize the results obtained in Sec. \ref{sec:summary},
especially, stressing the applicability of our formalism to 
searches in the recently found second-generation TDI data 
combinations.\cite{Cornish:2003tz,Shaddock:2003dj,Tinto:2003vj} 
Note that in the expressions appearing in this paper, we set the 
gravitational coupling constant ($G$) and the speed of light 
($c$) to unity.

\section{\label{sec:pseudo}The pseudo-detectors}

As illustrated in Fig. \ref{LISAtri}, LISA consists of three
spacecrafts, labeled $i=$1, 2, and 3, located clockwise at the
correspondingly labeled vertices of an almost equilateral
triangle. Let the arm-lengths of this triangle be $L_1$, $L_2$,
and $L_3$, such that $L_i$ is the length of the arm facing the
vertex $i$. For specifying the orientation of each arm, we assign
unit vectors $\hat{\bf n}_i$ along them, such that their directions
are oriented anticlockwise about the triangle. Each spacecraft
will have two optical benches (denoted by $i$ and $i^*$ in Fig.
\ref{LISAtri}) equipped with independent lasers
and photodetectors. Thus, each spacecraft will shoot two beams
towards the other two spacecrafts, respectively, resulting in six
one-way beams along the 3 arms.

The effect of an impinging GW
is to cause a shift in
the frequency of the laser beams. But a GW is not the only source causing
such a shift. The laser-frequency fluctuation is another source.
If $\nu_0$ is the central frequency
of all the lasers in LISA, then the fractional shift caused by such
fluctuations in the beam originating at optical bench $i$ is
\be C_i (t) \equiv
\frac{\D\nu_i(t)}{\nu_0} \ \ . \ee
Similarly for the beam from the bench $i^*$.
What is measured, however, is the frequency fluctuation in the
beam from one bench relative to that in the beam from a bench
in one of the two other vertices. This way one obtains three ``clockwise''
oriented streams, $U_i$, and three ``anticlockwise'' oriented streams,
$-V_i$. The frequency fluctuation in the beam from
bench $(i-1)$ relative to that in the beam originating at bench $i^*$ is
termed $U_i$.
\footnote{Note that the indices $i$ and $i\pm 1$ can take only 1, 2,
and 3 as values. These three numbers are ordered clockwise in Fig.
\ref{LISAtri}. By convention, whereas $i+1$ equals the number next to
$i$ while going clockwise in that figure, $i-1$ equals the number
preceding $i$. E.g., when $i=3$, we take $i-1=2$ and $i+1=1$; when
$i=1$, we take $i-1=3$ and $i+1=2$.}
Therefore, $U_1(t) \equiv C_3(t-L_2)-C_1(t)$ is the data stream
measured by beating the beam transmitted by bench 3
against that of bench $1^*$, measured at time $t$ at bench 1.
The remaining two streams, namely, $U_2$ and $U_3$,
can be obtained by cyclic permutation of the
indices in the $U_1$ expression.
(Thus, $U_2(t) \equiv C_1(t-L_3)-C_2(t)$, and so on.)
Three more data streams, termed $-V_i$,
are obtained by measuring the frequency fluctuation in the
beam from bench $(i+1)^*$ relative to that originating at bench $i$.
For instance, $-V_1(t) \equiv C_2(t-L_3) -C_1(t)$. Once again, the
remaining two $-V_i$ streams can be obtained by cyclic permutation of the
indices in the above expression for $-V_1$.
(Hence, $-V_2(t) \equiv C_3(t-L_1) -C_2(t)$, and so on.)

The fractional shift $C_3(t-L_2)$
is constructed from $C_3(t)$ by shifting back the latter stream
in time by an amount $L_2$.
For brevity of expressions, we introduce the time-shift operator $\zeta_i$
by its action on a data stream $x(t)$ as in:
\be \label{operator}
\zeta_i x(t)= x(t-L_i) \ \ ,
\ee
where the label $i$ denotes the arm along which
the time-shift is affected. One can thus define the 6 inter-craft streams as
follows\cite{Dhurandhar:2001kx}:
 \bea \label{datastreams} U_1 &=&
\zeta_2C_3 -C_1\,,\quad U_2=\zeta_3C_1-C_2\,,\quad
U_3=\zeta_1C_2-C_3 \ \ ,\no\\
V_1 &=& C_1 -\zeta_3C_2\,,\quad V_2=C_2-\zeta_1C_3\,,
\quad V_3=C_3-\zeta_2C_1 \,.
\eea
Note that the effect of such a shift on the
Fourier components, $\tilde{x}(f) \equiv \int_{-\infty}^\infty
x(t)e^{2\pi {\rm i} ft}dt$, of the data is to change them to
$e^{-2\pi{\rm i}f L_i}\tilde{x}(f)$, where $f$ is a frequency variable.
Therefore, the effect of the operator $\prod_{i=1}^3 \zeta_i^{\a_i}$ on
a data stream is to change its Fourier transform by the factor
$e^{-2\pi{\rm i}f \sum_{i=1}^3\a_iL_i}$. 

These data streams, however, are expected to suffer from several
noise sources, two of which, viz., the laser-frequency fluctuations and
the optical-bench motions, tower over the others.
The two other noise sources are the photon-shot noise 
and the fluctuations in the motion of the proof masses.
The laser-frequency fluctuations directly influence the fractional
frequency-shifts, $C_i$, and, therefore, the data streams $U_i$ and $V_i$.
The effect of the optical bench motions on the $C_i$ is additive:
Let the random velocities
of the optical benches be ${\bf v}_i$ and ${\bf v}_{i^*}$. This causes
a Doppler shift in the frequency of the lasers mounted on those benches,
which in turn modifies the $C_i$ to
\bea
C_1 &\to& C_1 - \hat{\bf n}_3\cdot {\bf v}_1 \ \ ,\no \\
C_{1^*} &\to& C_{1^*} + \hat{\bf n}_2\cdot {\bf v}_{1^*} \,.
\eea
The remaining $C_i$ and $C_{i^*}$ can be inferred by cyclically permuting
the indices in the above expression. If ${\bf u}_{i,i^*}$ are the
random velocities of the proof masses on benches $i$ and $i^*$, then it is
easy to see that the value of $U_1$ and $V_1$ gets affected by additional
terms $2{\bf \hat n}_2 \cdot {\bf u}_{1^*}$ and
$2{\bf \hat n}_3 \cdot {\bf u}_1$, respectively.
Finally, when there
is a GW signal present, these streams will receive additional contribution
owing to the fractional frequency shifts caused by it. Thus, in the
presence of the above noise sources and signal, the data streams
get modified as
\bea \label{intercraft}
U_1 &\to& U_1 = \zeta_2(C_3 - \hat{\bf n}_2\cdot {\bf v}_{3})
-(C_{1^*} + \hat{\bf n}_2\cdot {\bf v}_{1^*})
+2{\bf \hat n}_2 \cdot {\bf u}_{1^*}
+n^{\rm shot}_{U_1} +U_1^{\rm GW}\ \ ,\no \\
V_1 &\to& V_1 =-\zeta_3(C_{2^*} + \hat{\bf n}_3\cdot {\bf v}_{2^*})
+(C_1-\hat{\bf n}_3\cdot {\bf v}_{1}) +2{\bf \hat n}_3 \cdot {\bf u}_1
-n^{\rm shot}_{V_1} - V_1^{\rm GW} \ \ , \eea
where $n^{\rm
shot}_{U_i,V_i}$ are the photon-shot noises and $U_i^{\rm GW}$ and
$V_i^{\rm GW}$ are the GW signals present in the $U_i$ and $V_i$
data streams, respectively. The four remaining $U_i$ and $V_i$
streams can be obtained from the above expressions by cyclic
permutations of the indices. In the rest of the paper the $U_i$ and
$V_i$ will refer to these modified streams. When a (GW) signal is
absent, they will still be given by these modified expression, but
with $U_i^{\rm GW}=0$ and $V_i^{\rm GW}=0$. The form of $U_i^{\rm
GW}$ and $V_i^{\rm GW}$ in the presence of a signal will be explored
in the next section.

In addition to the six inter-craft data streams discussed above there
is supplementary information available about the noise sources in the
intra-craft beams exchanged through the optical fibers connecting two
adjacent optical benches $i$ and $i^*$ located in the $i$th craft.
In all there are six intra-craft beams, two per bench pair. But the two
intra-craft beams within a craft can be beaten against each other to produce
a single stream that is 
directly relevant to noise suppression. To wit, at craft 1, by beating
the frequency of the beam from bench 1 relative to that on bench $1^*$,
one forms the stream,
\be\label{intracraft}
W_1 = \left(C_1-\hat{\bf n}_3\cdot {\bf v}_1\right)
- \left(C_{1^*}+\hat{\bf n}_2\cdot {\bf v}_{1^*}\right)
+ \hat{\bf n}_3\cdot {\bf u}_1 + \hat{\bf n}_2\cdot {\bf u}_{1^*} \,.
\ee
Two other intra-craft data combinations, $W_2$ and $W_3$, can be
obtained by the cyclic permutation of indices in the above expression.
Note that these intra-craft streams will bear negligible influence
from any impinging gravitational wave. Nevertheless, as we explain below,
they offer
information on laser-frequency fluctuations and optical-bench motions
that can be used to render the $U_i$ and $V_i$ streams essentially free
of any noise from these two sources.
Together with the $U_i$ and $V_i$, the $W_i$ form a a total of nine
data streams that a data analyst has recourse to in hunting for a GW
signal in LISA.

Following the work of Tinto and Armstrong \cite{Tinto:yr}, it was
shown by Dhurandhar et al. \cite{Dhurandhar:2001kx}
that by acting on the 6 inter-craft
streams, $U_i$, $V_i$, and the 3 intra-craft streams, $W_i$, with
certain polynomials, $p^A_i$, $q^A_i$, and $r^A_i$, of the time-shift
operators, $\zeta_i$, one can form several combinations of time-delayed
data streams,
\be
\label{pseudoGen} x^{A}(t) = \sum_{i=1}^{3}\left[p^A_iV_i(t) +
q^A_iU_i(t) + r^A_iW_i(t)\right] \ \ ,
\ee
that have the laser-frequency and the optical-bench motion noise
eliminated. Above, $A$ labels the different combinations so obtained.
The above technique of constructing such data combinations of
pseudo-detectors is called time-delay interferometry.
The pseudo-detectors can be recast as
\be
x^{A} = {\rm Trace}\left[{\bf e}^A\cdot {\bf Z}\right] \ \ ,
\ee
where
\be
{\small
{\bf e}^A = \left(\begin{array}{ccc}
p^A_1 & p^A_2 & p^A_3 \\
q^A_1 & q^A_2 & q^A_3 \\
r^A_1 & r^A_2 & r^A_3
\end{array}\right) }
\quad{\rm and}\quad
{\small
{\bf Z} = \left(\begin{array}{ccc}
V_1 & U_1  & W_1 \\
V_2 & U_2  & W_2 \\
V_3 & U_3  & W_3 \\
\end{array}\right) \,.}
\ee
Thus, for a given choice of the data streams $U_i$, $V_i$, $W_i$ (and,
therefore, the matrix ${\bf Z}$), the matrix ${\bf e}^A$ of the
time-shift polynomials forms a representation of the pseudo-detectors
$x^A$.

Of the several possible pseudo-detectors, only 3 are linearly
independent and have a non-vanishing GW strain in them, in general
\cite{Prince:2002hp,Nayak:2003na}.
The three that will be discussed here are defined by their corresponding
${\bf e}^A$:
\be
{\small
{\bf e}^1 = \left(\begin{array}{ccc}
1-\zeta & 1+2\zeta & -2-\zeta \\
1+2\zeta & 1-\zeta & -2-\zeta \\
\zeta^2-1 & \zeta^2-1 & 2(1-\zeta^2) \\
\end{array}\right) \>,}
\quad
{\small
{\bf e}^2 = \left(\begin{array}{ccc}
-\zeta-1 & 1 & \zeta \\
-1 & 1+\zeta & -\zeta \\
1-\zeta^2 & -1+\zeta^2 & 0 \\
\end{array}\right) \>,}
\ee
and
\be
{\small
{\bf e}^3 = \left(\begin{array}{ccc}
1 & 1 & 1 \\
1 & 1 & 1 \\
-1-\zeta & -1-\zeta & -1-\zeta \\
\end{array}\right) \,.}
\ee
In the above expressions, it is assumed that all arm lengths
are almost identical. Therefore, $\zeta_1 \simeq \zeta_2 \simeq
\zeta_3 \equiv \zeta$. It is important to note that these data
combinations diagonalize their noise covariance matrix
\cite{Nayak:2002ir} and, therefore, are also noise-independent.

Although the laser-frequency noise is eliminated in the $x^{A}(t)$,
there is still present the
noise associated with the acceleration of the proof
masses onboard each craft and the photon-shot noise. In general,
\be
x^{A}(t) = n^A(t) + h^A(t) \ \ ,
\ee
where
\be\label{strainTime}
h^A(t) =  \sum_{i=1}^{3}\left[p^A_iV_i^{\rm GW}(t)
+q^A_iU_i^{\rm GW}(t)\right] \ \ ,
\ee
is the gravitational-wave strain in pseudo-detector $A$ and $n^A(t)$
is the time-delayed sum of the remaining noise components.
One typically assumes that these components
and, hence, the total noise, has a Gaussian probability distribution with
a zero mean. Their variance-covariance matrix elements are given as follows.
For the photon-shot noise, these elements are
\bea\label{shotPSD}
\overline{\tilde{n}^{{\rm shot}*}_{U_i}(f)~\tilde{n}^{\rm shot}_{U_j}(f')}
&=& \overline{\tilde{n}^{{\rm shot}*}_{V_i}(f)~\tilde{n}^{\rm shot}_{V_j}(f')}\no\\
&=& \overline{\tilde{n}^{{\rm shot}*}_{W_i}(f)~\tilde{n}^{\rm shot}_{W_j}(f')}= \frac{1}{2}P^{\rm shot}(f)\delta(f-f')\delta_{ij} \>,\no\\
\overline{\tilde{n}^{{\rm shot}*}_{U_i}(f)~\tilde{n}^{\rm shot}_{V_j}(f')}
&=&\overline{\tilde{n}^{{\rm shot}*}_{V_i}(f)~\tilde{n}^{\rm shot}_{W_j}(f')}
=\overline{\tilde{n}^{{\rm shot}*}_{W_i}(f)~\tilde{n}^{\rm shot}_{U_j}(f')}
= 0 \ \ ,
\eea
for any $i$ and $j$. Above, $P^{\rm shot}(f)$ is termed as the one-sided
power-spectral density (PSD) of the photon-shot noise.
Furthermore, one assumes the proof-mass noise
to be isotropic, such that:
\bea\label{proofPSD}
\overline{\left(\hat{\bf n}_i\cdot\tilde{\bf u}_j^*(f)\right)
~\left(\hat{\bf n}_k\cdot\tilde{\bf u}_l(f')\right)}
&=& \overline{\left(\hat{\bf n}_i\cdot\tilde{\bf u}_{j^*}^*(f)\right)
~\left(\hat{\bf n}_k\cdot\tilde{\bf u}_{l^*}(f')\right)}
=\frac{1}{2}P^{\rm proof}\delta(f-f')\delta_{jl} \>,\no\\
\overline{\left(\hat{\bf n}_i\cdot\tilde{\bf u}_{j^*}^*(f)\right)
~\left(\hat{\bf n}_k\cdot\tilde{\bf u}_l(f')\right)}
&=&\overline{\left(\hat{\bf n}_i\cdot\tilde{\bf u}_j^*(f)\right)
~\left(\hat{\bf n}_k\cdot\tilde{\bf u}_{l^*}(f')\right)} =0 \ \ ,
\eea
for any $i$, $j$, $k$, and $l$. Also, the covariance of the shot noise in
any data stream with the noise in the motion of any proof-mass is zero.
It is estimated that
$P^{\rm{shot}}=1.8 \cross 10^{-37}[{f/1 \rm{Hz}}]^2 {\rm Hz}^{-1}$
and that the proof-mass noise PSD is $P^{\rm{proof}}=2.5 \cross
10^{-48}[{f/1\rm{Hz}}]^{-2} {\rm Hz}^{-1}$.\cite{SysTech:2000rep}
While the proof-mass noise enters the very low
frequency band of LISA, the shot noise enters the higher end of
LISA's sensitivity band.

It is now possible to deduce the noise PSD, $P^{(A)}(f)$, of each of
the pseudo-detectors from the above expressions. It follows from them
that in the absence of
a signal, each pseudo-detector is pure noise, $x^A (t)\equiv n^A(t)$, with
a zero mean Gaussian probability distribution. The variance of this
distribution is
\be
\overline{\tilde{n}^{A*}(f)\tilde{n}^{B}(f')}
=\frac{1}{2}P^{(A)}(f)\delta(f-f')\delta^{AB} \,.
\ee
By substituting for $x^A$ from Eq. (\ref{pseudoGen}) (with $U^{\rm GW}_i$
and $V^{\rm GW}_i$ set to zero there) in the above equation and using the
covariances of the noise components defined in Eqs (\ref{shotPSD})
and (\ref{proofPSD}) one finds
\cite{Dhurandhar:2001kx}:
\bea \label{generalNoise}
P^{(A)}(f) &=& \sum_{i=1}^3 \Big[(|2p_i^A + r_i^A|^2 +
|2q_i^A + r_i^A|^2)P^{\rm{proof}} \no\\
&& \quad\quad+(|2p_i^A|^2 + |2q_i^A|^2)P^{\rm{shot}}\Big] \eea The
resulting noise spectra for each pseudo-detector is
\cite{Krolak:2004xp}: \bea \label{noise} P^{(1)}(f) &=& P^{(2)}(f) =
8\sin^2 (\pi f L)\big\{\left[2+\cos(2\pi f L)\right] P^{\rm{shot}} \no\\
&&\quad \quad \quad\quad
+[6 + 4\cos(2\pi f L) + 2\cos(4\pi f L)]P^{\rm{proof}}\big\}\ \ ,\no\\
P^{(3)}(f) &=& \left[2+4\cos(2\pi f L)\right]\left[P^{\rm{shot}} +
4\sin^2(\pi f L)P^{\rm{proof}}\right] \,. \eea Therefore, the data
analysis challenge is to detect signals in this remaining noise.

\section{\label{sec:gwstrain}The Signal}

\begin{figure}[!hbt]
\centerline{\psfig{file=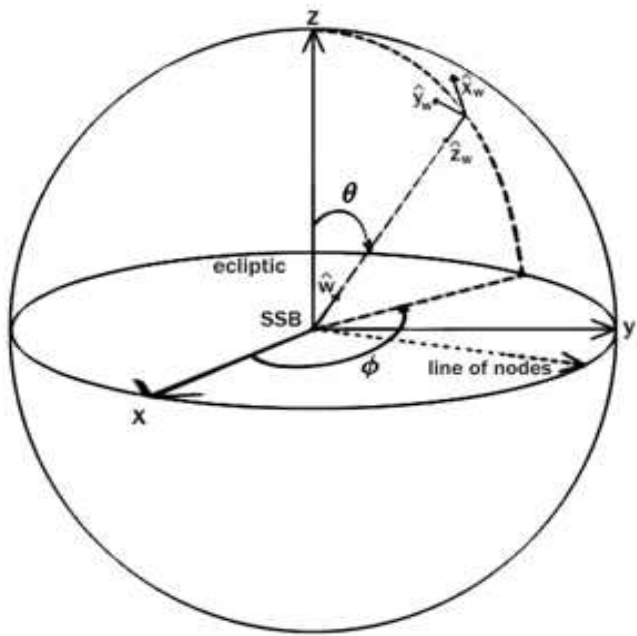,height=3.5in,width=3.5in}}
\caption{The solar-system barycentric (SSB) frame, denoted by the
$(x,y,z)$ axes. The angles $(\theta,\phi)$ specify the sky position,
$\hat{\bf w}$ of a
GW source. The axes of the wave-frame associated with a gravitational
wave emanating from that source are labeled by the unit-normal vectors
$(\hat{\bf x}_w,\hat{\bf y}_w,\hat{\bf z}_w)$.
The orientation of the wave frame relative to the SSB frame is given by
the Euler angles, $\left(\phi-\pi/2,\pi-\theta, \psi\right)$, where $\psi$
is the angle between the line of nodes and $\hat{\bf x}_w$ (as explicitized
further on the left panel in Fig. \ref{sourcewave}). The celestial longitude is
drawn as a dashed arc passing through the origin of the wave frame.
} \label{barycenter}
\end{figure}

Since LISA will be orbiting the solar-system barycenter (SSB), it is
convenient to introduce a reference frame centered at the SSB. 
As shown in Fig. \ref{barycenter}, we
define the $\hat{\bf x}$ and $\hat{\bf y}$ axes of this SSB frame
to lie on the ecliptic, and the  $\hat{\bf z}$ axis to be normal to it
and pointing towards the north ecliptic pole. The $\hat{\bf x}$ axis
points towards the vernal equinox. We take the GW source to be located 
in the direction given by the vector
$\hat{\bf w}$. A gravitational wave from this source will arrive at the SSB
origin traveling along $-\hat{\bf w}$. 
The sky position $\{\theta,\phi\}$ defines the
Cartesian components of the propagation direction, i.e., $\hat{\bf
w} = (\sin\theta\cos\phi,\sin\theta\sin\phi,\cos\theta)$. Thus, the
sky-position angles are equivalently characterized by two of the
three components of $\hat{\bf w}$, say, $w_1=\sin\theta\cos\phi$ 
and $w_2=\sin\theta\sin\phi$. Also, $\theta_k$ and $\phi_k$ define 
the plane transverse to $\hat{\bf
w}$: \be \hat{\mbox{\boldmath$\theta$}} \equiv \frac{\partial
{\hat{\bf w}}}{\partial \theta}\ \ ,\quad
\hat{\mbox{\boldmath$\phi$}} \equiv \frac{1}{\sin\theta}
\frac{\partial {\hat{\bf w}}}{\partial \phi}\,. \ee

The perturbation created by the wave at a spacetime location 
$(t, {\bf r})$ is given by
\be \label{metricPerturbation} h_{kl}(t,
{\bf r})=h_+(t-\hat{\bf w} \cdot {\bf r})(\theta_k \theta_l - \phi_k
\phi_l)+ h_\cross (t-\hat{\bf w} \cdot {\bf  r})( \theta_k \phi_k +
\theta_l \phi_l) \ \ , \ee where $h_+(t)$ and $h_\cross(t)$ are the
two GW polarizations and ${\bf r}$ is the position vector of the 
spatial location of the perturbation in the SSB frame.
In the time domain, the strain induced along the $i$th arm is
\be
\label{timedomainSignal} h_i(t) = h_{kl}(t)n_i^k n_i^l =
h_{i+}(t)\xi_{i+}(w_1,w_2) +h_{i\cross}(t)\xi_{i\cross}(w_1,w_2)\ \ ,
\ee
where we used the Einstein summation convention over the repeated
indices $k$ and $l$. Above, \be \label{directionVecs} \xi_{i+} =
(\hat{\mbox{\boldmath$\theta$}} \cdot {\hat{\bf n}}_i)^2 - (\hat
{\mbox{\boldmath$\phi$}} \cdot {\hat{\bf n}}_i)^2 \ \ , \quad
\xi_{i\cross} = 2(\hat{\mbox{\boldmath$\theta$}} \cdot {\hat{\bf
n}}_i) (\hat{\mbox{\boldmath$\phi$}} \cdot {\hat{\bf n}}_i) \ee are
the beam-pattern functions of the $i$th arm for the two
polarizations.

An impinging GW causes a change in the light-travel time along an
arm that can be calculated by solving the null geodesic equation
in the corresponding perturbed spacetime. This in turn causes a
time-varying Doppler shift, which clearly depends on the
difference between the GW strains at the two space-crafts at the end of the
arm. One also expects this shift to be dependent on the position
of the source relative to the arm, ${\hat{\bf w}}\cdot{\hat{\bf n}}_i$.
Thus, the GW contribution to the data stream $V_i$ is
given by \cite{Dhurandhar:2001kx}
\be \label{fractionShift}
V_i^{\rm GW}(t) = {-1\over 2(1-\hat{\bf w} \cdot {\hat
{\bf n}}_i)} \left [h(t-\hat{\bf w} \cdot {\bf  r}_{i+1})-h(t-\hat{\bf w}
\cdot {\bf r}_{i-1}-L) \right] \ee
where ${\bf r}_{i}$ is the position vector
of the $i$th craft in the LISA frame.

\begin{figure}[!hbt]
\centerline{\psfig{file=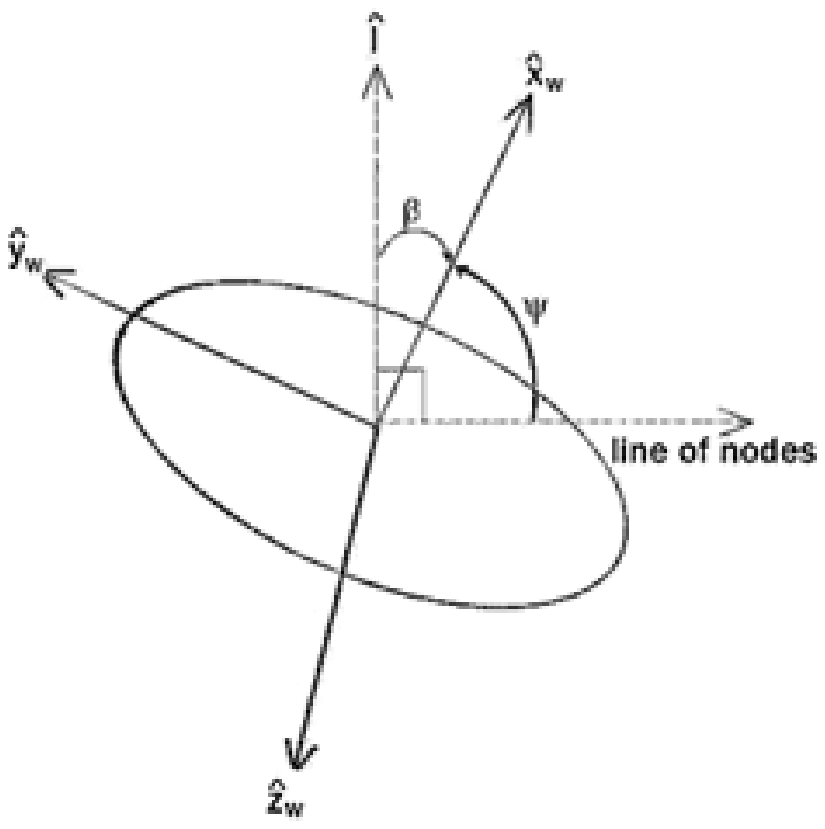,height=3.in,width=3.in}
\psfig{file=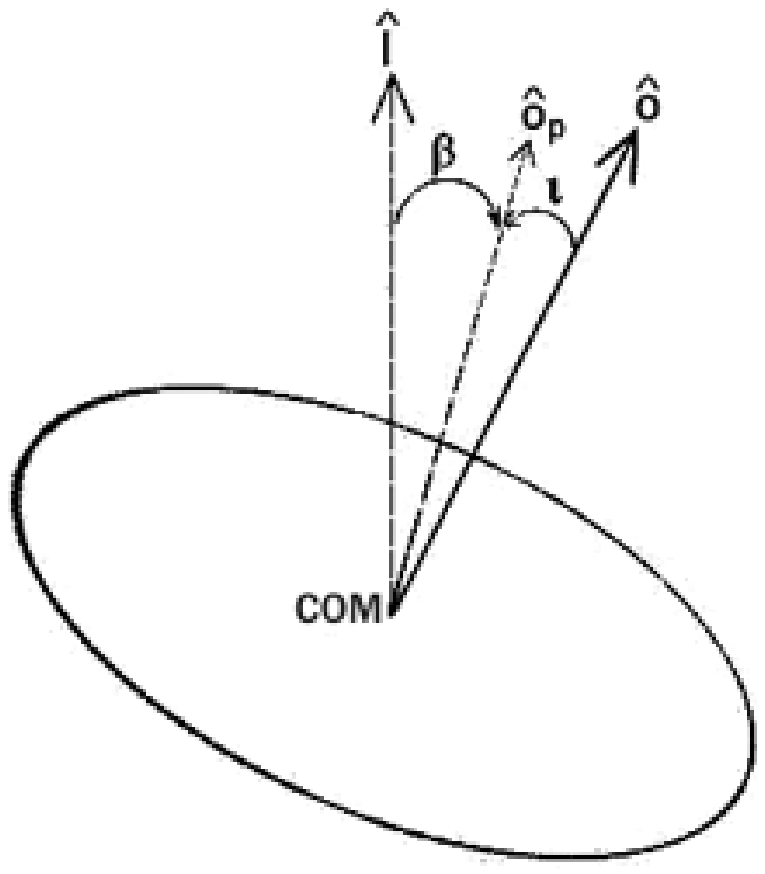,height=3.in,width=3.in}}
\caption{The left panel shows the wave frame and the right panel shows
the orientation of the compact binary's orbit. The orientation of the wave
frame is such that its $\hat{\bf z}_w$ axis points towards the origin of
the SSB frame (as shown in Fig. \ref{barycenter}) and its  $\hat{\bf x}_w$
axis lies along the semi-major axis of the wave's polarization ellipse.
The tangent, $\hat{\bf l}$, to the longitude at the source's sky-position
is perpendicular to the line of nodes, which lies on the ecliptic.
The two form a plane on which $\hat{\bf x}_w$ lies
making an angle $\psi$ with the line of nodes. Therefore, $\beta =
\pi/2 -\psi$. In the right panel, COM is the center of mass of the binary.
The normal, $\hat{\bf o}$, to its orbit is along the binary's orbital angular
momentum vector and has an inclination of angle $\epsilon$
(introduced in the main text) with the line of sight, $\hat{\bf w}$.
The projection of $\hat{\bf o}$ on the celestial sphere is $\hat{\bf o}_p$,
which makes an angle $\beta$ with respect to $\hat{\bf l}$. The angle between
$\hat{\bf o}$ and $\hat{\bf o}_p$ is $\iota$. These two angles,$\beta$
and $\iota$, completely specify the orientation of the orbit's normal. Thus,
$\iota = \pi/2 -\epsilon$.
}
\label{sourcewave}
\end{figure}

We now consider the effect of a signal from a non-spinning
compact binary, with member masses $m_1$ and $m_2$, on 
$V_i^{\rm GW}(t)$. The two polarization amplitudes for the $i$th arm are
\bea
\label{GWstrains} h_{i+\hspace{1pt}}(t) = H(\Omega_i)
\left[{1+\cos ^{2}\epsilon\over 2} \cos 2\psi \cos (\Phi_i (t)+
\delta_c) + \cos \epsilon \sin 2\psi \sin (\Phi_i(t) + \delta_c) \right]
&&\ \ ,
\label{GWstrains+}\\
h_{i\cross \hspace{1pt}}(t) = H(\Omega_i) \left[-{1+\cos
^{2}\epsilon\over2 }\sin 2 \psi \cos (\Phi_i (t)+\d_c) + \cos
\epsilon \cos 2 \psi \sin (\Phi_i (t) + \d_c) \right] &&\ \ ,
\label{GWstrainsx} \eea
where $\Phi_i (t)$ is the phase received at that arm, $\{\psi,\epsilon\}$
are the polarization and inclination angles of the binary source
(as explained in Fig. \ref{sourcewave}) and $\delta_c$ is the initial phase at
the origin of the LISA frame. For binaries that include 
a white-dwarf, the signal will be essentially monochromatic, with  
$\Phi_i (t) = \Omega_i t$. Above, $H(\Omega_i)$ is the signal
amplitude at LISA defined as:
\be \label{amplitude}
H(\Omega_i)=1.188\cross 10^{-22} \left[{\mathcal{M}\over
1000~M_\odot} \right]^{5/3} \left[{R\over 1{\rm Gpc}}\right]^{-1}
\left[{\Omega_i \over 2 \pi \hspace{2pt}\times 1{\rm mHz}}\right]^{2/3}
\ee
where $\mathcal{M} = (m_1m_2)^{3/5}/(m_1+m_2)^{1/5}$
is the chirp mass, $R$ is the distance to the source and
\be \label{omega}
\Omega_i=\Omega_0 \gamma_i \ee
is the Doppler
shifted source frequency, $\Omega_0$, at the $i$th arm owing to
LISA's motion with respect to the solar-system barycenter.
Note that $\gamma_i$ is the sky-position
dependent Doppler factor,
\be\label{gamma}
\gamma_i = \left(1-\hat{\bf w}\cdot\hat{\mbox{\boldmath$\eta$}}_i\right) \ \ ,
\ee
where $\hat{\mbox{\boldmath$\eta$}}_i$ 
is the velocity of the geometrical center of
the $i$th arm. Substituting the GW strain deduced from
Eqs. (\ref{GWstrains}) and (\ref{timedomainSignal}) in the
$V_i^{\rm GW}(t)$ expression (\ref{fractionShift}), mainfests its
dependence
on the source parameters. One can similarly
obtain $U_i^{\rm GW}(t)$ as a function of these parameters.
We first discuss the case of the monochromatic signal. The extension
to slightly chirping sources is straightforward and will be presented
subsequently.

If the laser-frequency and optical-bench motion noises were at the level
of the other noises, one could hunt for GW signals in the $U_i$ and
$V_i$ data streams. But as we saw in the last section, this is not the case
and, hence, one has to search for them in the pseudo-detectors $x^A(t)$ in
which the contributions of these noises stand canceled. Working with
the $x^A(t)$, however, makes the data analysis formulation a little
non-trivial since one has to contend with time-delaying appropriately
the six inter-craft data streams that can potentially harbor a GW signal,
$h^A(t)$. Implementing this is somewhat easier in the Fourier domain.
Thus, the algorithm we follow in the rest of this section is as follows.
We begin by first computing the Fourier transform of the data streams
${V}_i^{\rm GW}$ (and ${U}_i^{\rm GW}$). These will then be time-delayed
and combined to calculate the frequency components $\tilde{h}^A(\omega)$
(where $\omega = 2\pi f$) of the GW signal in the $A$th pseudo-detector,
along the lines of Eq. (\ref{strainTime}). Its inverse Fourier transform
will finally yield $h^A(t)$,
which is the quantity we aim to search for in the $x^A(t)$. In the process,
we get an explicit expression for $\tilde{h}^A(\omega)$, which is useful
since the implementation of a search is faster in the Fourier domain, where
one can avail of the existing Fast Fourier transform
algorithms.\cite{numRecipes}

We begin by defining two new functions of $\{\psi,\epsilon\}$ that
appear naturally in the Fourier transforms, $\tilde{V}_i^{\rm GW}(\omega)$
and $\tilde{U}_i^{\rm GW}(\omega)$, of the six streams:
\bea
l_\cross &=& -{\rm i}\left(T_2^{2}(\psi, \epsilon, 0)
-T_2^{-2}
(\psi, \epsilon, 0) \right) \ \ ,\no \\
l_+ &=& \left(T_2^{2}(\psi, \epsilon, 0) + T_2^{-2}(\psi, \epsilon,
0) \right) \ \ , \eea
where $T_2^{~\pm 2}$ are  Gel'fand functions
\cite{Pai:2000zt}, \be T_2^{~\pm 2}\left(\psi,\epsilon,0\right) =
\frac{1}{4}\left(1\pm \cos\epsilon\right)^2 \exp\left(\mp {\rm
i}2\psi\right) \,.\ee
To find $\tilde{V}_i^{\rm GW}(\omega)$, we use the GW strain Eqs.
(\ref{GWstrains}) and (\ref{timedomainSignal}) in the
expression for $V_i^{\rm GW}(t)$ in Eq. (\ref{fractionShift}). Taking the
Fourier transform of the result gives
\bea \label{template}
\tilde{V}_i^{\rm GW} (\omega)=H(\Omega_0)T\Bigg[&&{\rm
sinc}\left(\left(\omega - \Omega_i\right)T\right)e^{-{\rm i}\delta_c} \times\no\\
&&\left(l_+^*(\psi, \epsilon) F_{V_{i+1};+}^*(\Omega_i) +
l_\cross ^*(\psi,
\epsilon) F_{V_{i+1};\cross}^*(\Omega_i )\right) \no\\
&+&{\rm sinc}\left(\left(\omega +\Omega_i\right)T\right) e^{{\rm i}\delta_c}\times\no\\
&&\left(l_+(\psi, \epsilon) F_{V_{i+1};+}(\Omega_i)
 +l_\cross(\psi, \epsilon) F_{V_{i+1};\cross}(\Omega_i)\right)
 \Bigg] \,.
\eea
The orientation of the $i$th arm resides in the functions,
\bea &&F_{V_{i+1};+, \cross}
=-{\rm i}b_{i} \hspace{1pt} \hspace{1pt} \xi_{{i};+, \cross}
\hspace{1pt} {\rm sinc} \left (\Omega_{i} L_{i} k_{i}^{-} \right)
e^{{\rm i}  \tau_{i}\Omega_{i} L_{i}} \ \ , \no\\
&&F_{U_{i-1};+, \cross} = {\rm i}b_{i} \hspace{1pt}
\hspace{1pt}\xi_{{i};+, \cross} \hspace{1pt} {\rm sinc} \left
(\Omega_{i} L_{i} k_{i}^+ \right) e^{{\rm i}  \tau_{i}\Omega_{i}
L_{i}} \ \ , \eea
where
\be \label{time Delay}
 b_i \equiv {\Omega_0 \gamma_i^{5/3} L_i \over 2}, \hspace{10pt}
  \hspace{10pt}   \tau_i \equiv{1 \over 2}
\left(1-{\hat{\bf w} \cdot \hat{\bf r}_i \over \sqrt 3 }\right ),
\hspace{10pt} k_i^\pm={(1\pm \hat{\bf w} \cdot \hat{\bf n}_i)\over 2} \ \ ,
\ee
are all real quantities. The $F_{U_{i};+, \cross}$ similarly determine the
fractional frequency shift $\tilde{U}_i^{\rm GW}$.
Note that the dependence on the angles $\{\psi, \epsilon\}$ has been
separated out in the form of $l_{+,\cross}$.
We will exploit this separation of variables in the next section to
eliminate the computational cost in searching over the $\{\psi, \epsilon\}$
angles for a GW signal.

The Fourier components of the GW strain $\tilde{h}^A(\omega)$
can now be found by combining the above $\tilde{V}_i^{\rm GW} (\omega)$
(and $\tilde{U}_i^{\rm GW} (\omega)$) via the Fourier analogue of
Eq. (\ref{strainTime}).
Thus,
\bea\label{strainA}
\tilde h^A(\omega) =  iH(\Omega_0)T\sum_{i}^{3} \Bigg[&&e^{-{\rm
i}(\delta_c+\sigma_i^A)} {\rm sinc}\left(\left(\omega -
\Omega_i\right)T\right) \hspace{3pt}
T_2^{\rho*}\hspace{3pt} D^{A*}_{\rho \hspace{1pt}i}e^{-{\rm i}  \tau_{i}\Omega_{i} L_{i}}\no\\
&& + e^{{\rm i}(\delta_c+\sigma_i^A)}{\rm sinc}\left(\left(\omega +
\Omega_i\right)T\right) T_2^{\rho}\hspace{3pt} D^{A}_{\rho
\hspace{1pt}i}e^{{\rm i}  \tau_{i}\Omega_{i} L_{i}} \Bigg] \ \ ,
\eea
where there is an implicit sum over $\rho=\pm 2$. Also, we
define
\be \label{dataCombinCompon}D^{A}_{\pm 2 \hspace{1pt}j}
\equiv b_j |M^{A}_{j}| \hspace{3pt} \left(\xi_{j_+} \mp {\rm i}
\xi_{j_\cross} \right) \ \ , \ee
where
\be M^A_{i} \equiv
q^A_{i-1}{\rm sinc} \left (\Omega_{i} L_{i} k_{i}^+ \right)
 -p^A_{i+1}{\rm sinc} \left (\Omega_i L_ik_{i}^-
  \right)
\ee
and $\sigma_i^A = \arg (M^A_{i})$.

The time-domain expression of the GW strain in the pseudo-detector
$A$ is obtained by taking the Fourier transform of $\tilde
h^A(\omega)$, and is found to be
\be\label{signalTime} h^A(t) =
H(\Omega_0) \sum_{j=1}^{3} \Re \left[e^{-{\rm i}\delta} E^{A*}_{j}
\hspace{3pt} S_j (t) \right] \ \ , \ee
where $\delta = \delta_c
+\pi/2$,
\be \label{signalComponents}
S_j^A (t) \equiv e^{{\rm i}\Omega_j (t -L_j\tau_j)+i\sigma_j^A} / g_j^A
\hspace{10pt} {\rm and} \hspace{10pt}
E^{A}_{j} \equiv g_j^A T^{~\rho}_2 \hspace{3pt}
D^{~A}_{j\hspace{1pt} \rho} \,. \ee
Above, $g_j^A$ is a
normalization constant such that
\be \langle S_j^A, S_j^A \rangle_{(A)}
\equiv 4\int_0^\infty df
\frac{\left|\tilde{S}_j^{A}(f)\right|^2} {P^{(A)}(f)}=1
\ \ , \ee
which implies that for an observation duration (i.e.,
signal integration time) $T$,
\be g_j^A = \left[\frac{2T}{\pi
P^{(A)}(\Omega_j)}\right]^{1/2} \,. \ee
When considering a slightly chirping source (i.e., when 
$\dot{\Omega}_0 \ll \Omega_0/T$) one can expand the signal phase as
\be\label{chirpPhase}
\Phi_i(t) = \Omega_i t + {1\over2}\dot{\Omega}_i t^2 \,.
\ee 
In that case, the only modification to $h^A(t)$ occurs in the $S^A_i$ term:
\be \label{signalChirping}
S_j^A (t) = e^{{\rm i}\Omega_j (t-L_j\tau_j)
+{\rm i}({1\over2}\dot{\Omega}_jt^2+\sigma_j^A)} / g_j^A \ \ ,
\ee
which defines the time-domain template for chirping compact-object binaries.
Note that apart from the normalization constant, the template 
$S_j^A(t)$ is a pure phase term. We will find this useful when deducing 
the matched filter in the following section.

Equations (\ref{omega}) and (\ref{gamma}) show that $\dot{\Omega}_j$ is 
related to the sky position and the intrinsic chirp rate $\dot{\Omega}_0$.
In order to ensure that the phase evolution of a template models that of a
signal well, it is important to allow for non-zero $\dot{\Omega}_0$ in 
searches of binaries involving masses higher than those of white dwarfs, such 
as in searches of neutron-star binaries and binaries involving a neutron star
and a black hole. This is because the gravitational radiation reaction on 
these sources is stronger than those involving white-dwarfs. The 
post$^1$-Newtonian waveform reveals that for small chirp masses and source 
frequencies, the waveform phase can be expanded as in 
Eq. (\ref{chirpPhase}), with
\be
\dot{\Omega}_0 = \frac{48}{5}\left(\frac{{\mathcal M}}{2}\right)^{5/3}
\Omega_0^{11/3} \ \ ,
\ee
Thus, determining $\dot{\Omega}_0$ is significant since, 
together with $\Omega_0$,
it determines the binary chirp mass ${\mathcal{M}}$. And as shown by 
Eq. (\ref{amplitude}), additional knowledge of the amplitude will then help 
in estimating the distance to the binary.\cite{Schutz}

To summarize, the GW strain in pseudo-detector $A$ is given by Eq.
(\ref{strainA}) and is determined by eight independent parameters, 
$\{R,\delta,\Omega_0,
\dot{\Omega}_0,\psi,\epsilon,\theta,\phi\}$. To search for a signal we
must devise a strategy to seek these strains for a range of parameter values
accessible to LISA's pseudo-detectors. This is what we deal with in the
next section. 

\section{\label{sec:statistic}The Optimal Statistic}

Given three independent pseudo-detectors, $x^A$, we now ask what is
the optimal detection statistic to look for GW signals, $h^A$, in
them. In the absence of any prior probabilities and costs, the
optimal detection strategy is the one that minimizes the rate of
false dismissals for a given rate of false alarms. This is termed as
the Neyman-Pearson criterion. Under this criterion, and for
zero-mean Gaussian noise, the detection statistic is the likelihood
ratio, $\lambda$, defined as \cite{Hels}
\be \label{logLR}
\ln\lambda = \sum_{A=1}^3 \left(\langle h^A, x^A \rangle_{(A)} - {1
\over 2} \langle h^A, h^A \rangle_{(A)} \right)\ \ , \ee
where the
first term is the sum of the cross-correlations of the expected
signal, $h^A$, with the respective data, $x^A$, over all
pseudo-detectors. The cross-correlation for pseudo-detector $A$ is
given by
\be
\langle h^A, x^A \rangle_{(A)} \equiv 4\Re\int_0^\infty
df\frac{\tilde{h}^{A*}(f)\tilde{x}^{A}(f)} {P^{(A)}(f)} \ \ ,
\ee
where $\Re(X)$ denotes the real part of a complex number $X$.
The second term in Eq. (\ref{logLR}) is an overall normalization
that is independent of the data. Substituting for $h^A$ from Eq.
(\ref{signalTime}), we get:
\be \label{matchedFilter} \sum_{A=1}^{3}
\langle h^A, x^A \rangle_{(A)} = \sum_{A=1}^{3} \sum_{i=1}^{3} \Re
\left[e^{-{\rm i}\delta} \hspace{3pt} E^{A*}_i \hspace{3pt}
C^{\hspace{1pt}A}_i \right] \ \ , \ee
where
\be \label{signal}
C^{A}_i \equiv \langle S_i^A, x^A\rangle_{(A)} \,. \ee
The double
summation in Eq. (\ref{matchedFilter}) can be replaced with the
single sum over a new index $k$,
\be \label{indicyconversion}
\sum_{A=1}^{3} \sum_{i=1}^{3} Y_i^A\equiv\sum_{k=1}^{9}
Y_{(k-1)\%3+1}^{\lceil {k \over 3} \rceil} \ \ , \ee
where $A = {\rm
Ceiling}(k/3) = \lceil {k \over 3} \rceil$ and $i={\rm Mod}(k-1,
3)+1 =(k-1)\%3+1$.\cite{ceilingMod} We simplify the above expressions
further by consistently identifying $Y_{(k-1)\%3+1}^{\lceil {k \over
3} \rceil}$ with $Y^k$, which form the components of a
9-dimensional vector ${\bf Y}$. Thus, $Y^{k=1} \equiv
Y^{A=1}_{i=1}$, $Y^{k=2} \equiv Y^{A=1}_{i=2}$, $Y^{k=3} \equiv
Y^{A=1}_{i=3}$, $Y^{k=4} \equiv Y^{A=2}_{i=1}$, and so on. We use
this algorithm to map the $D_i^A$, $E_i^A$, and $C_i^A$ to the
components, $D^k$, $E^k$, and $C^k$ of 9-dimensional vectors ${\bf
D}$, ${\bf E}$, and ${\bf C}$, respectively.

\begin{figure}[!hbt]
\centerline{\psfig{file=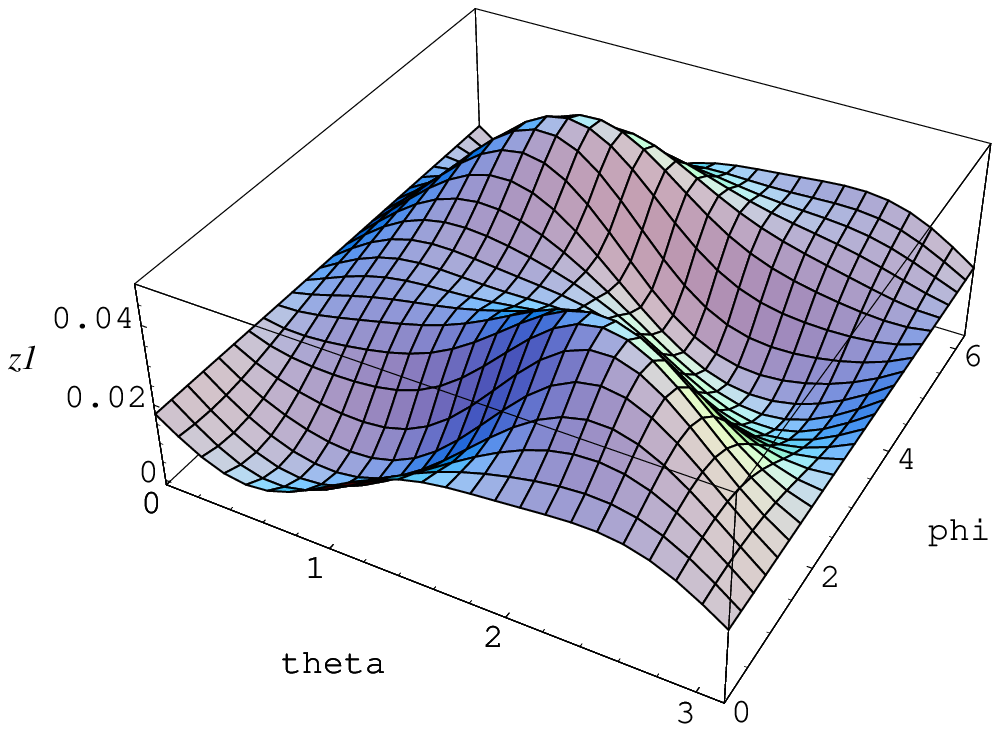,height=2.in,width=2.in}
\psfig{file=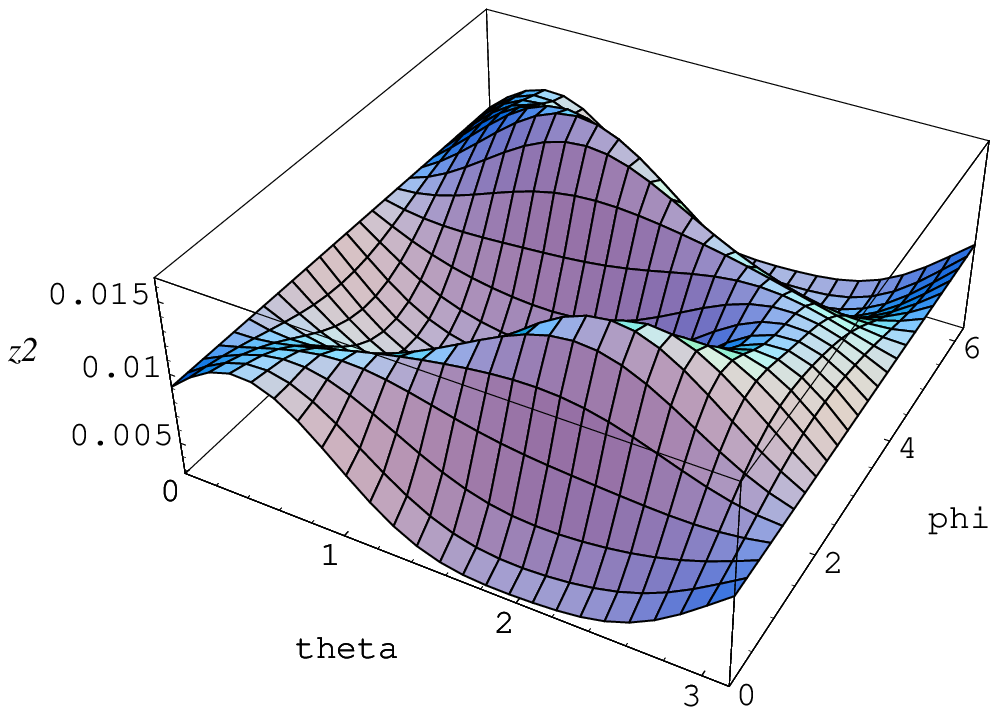,height=2.in,width=2.in}
\psfig{file=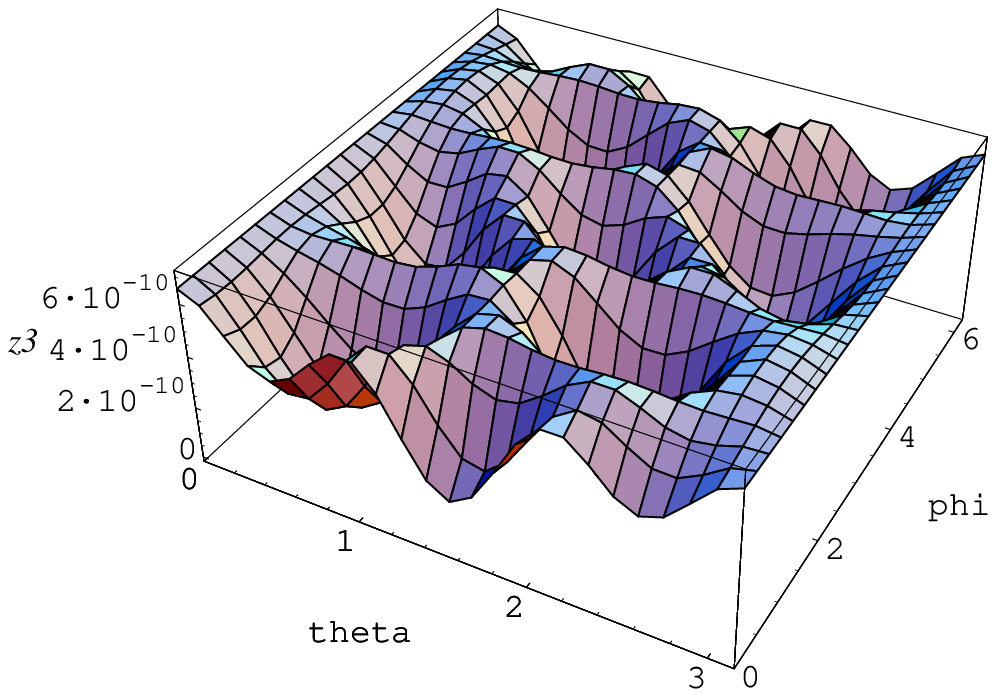,height=2.in,width=2.in}}
\caption{Sensitivities $z_A\equiv\sum_{i=1}^3 E_A^{i*}
E^A_i/(g_i^A)^2$ of the three pseudo-detectors as functions of sky
positions, $\{\theta,\phi\}$, in radians. Note that the Doppler
shift in the source frequency has been ignored here; so $g_i^A$ is
the same for all $i$, i.e., $g_i^A \equiv g^A$. These plots have
been evaluated for $\Omega_0=1$ mHz at the ``initial'' orbital
position of LISA labeled $t=0$. As illustrated above, direction of
maximum sensitivity varies from one pseudo-detector to another.}
\label{threeSens}
\end{figure}

\begin{figure}[!hbt]
\centerline{\psfig{file=Esquared1t0.eps,height=2.in,width=2.in}
\psfig{file=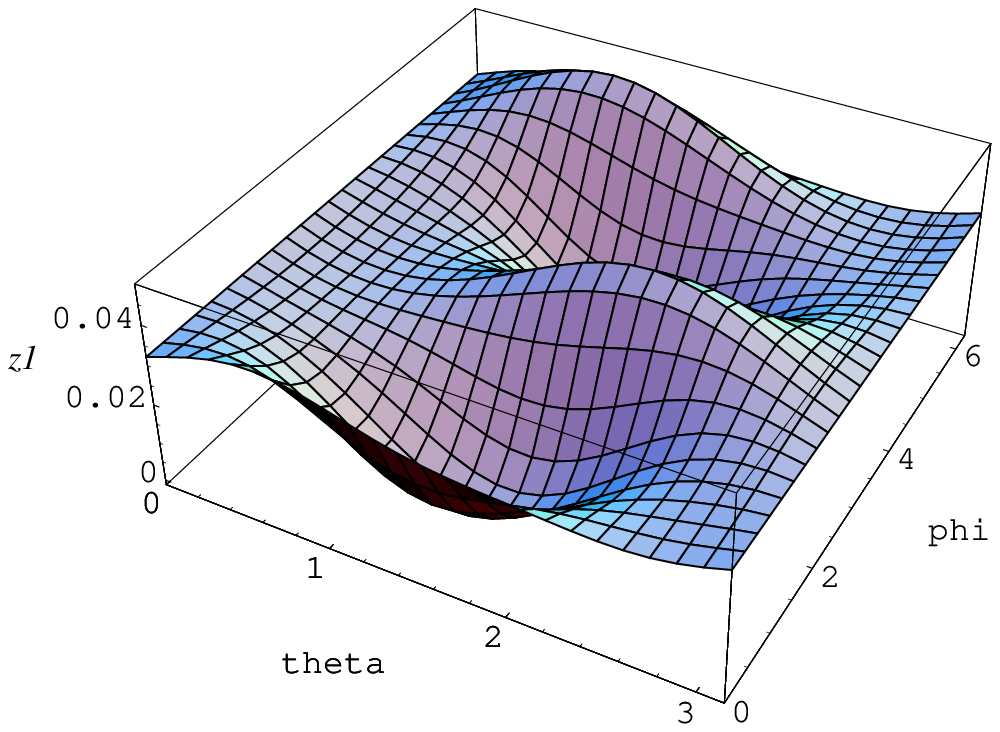,height=2.in,width=2.in}
\psfig{file=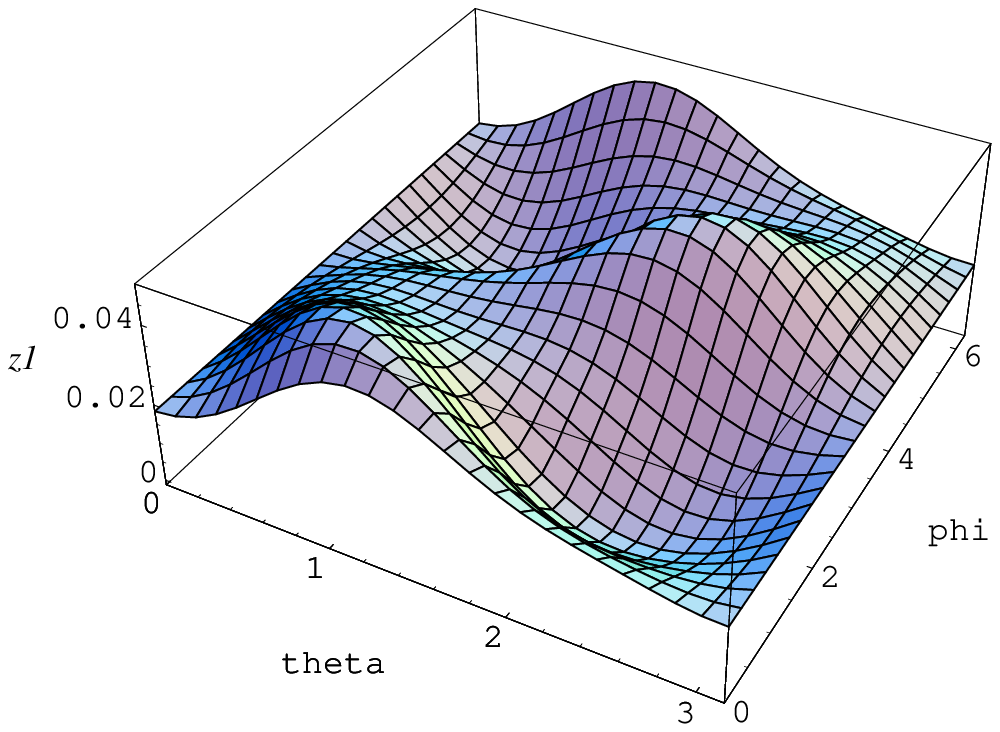,height=2.in,width=2.in}}
\caption{Sensitivity $z_1$, as defined in Fig. \ref{threeSens},
evaluated at $\Omega_0 =1$ mHz for three
different angular orbital positions (in radians), $\{0,{\pi\over3},{2\pi\over3}\}$,
with respect to the ``initial'' location denoted in Fig. \ref{threeSens}.
The left plot above is identical to
the left plot in Fig. \ref{threeSens} since it corresponds to the
same pseudo-detector and orbital location. Note
that the sky positions corresponding to the sensitivity maxima
vary from one location to another on LISA's orbit.}
\label{VarySens}
\end{figure}

\begin{figure}[!hbt]
\centerline{\psfig{file=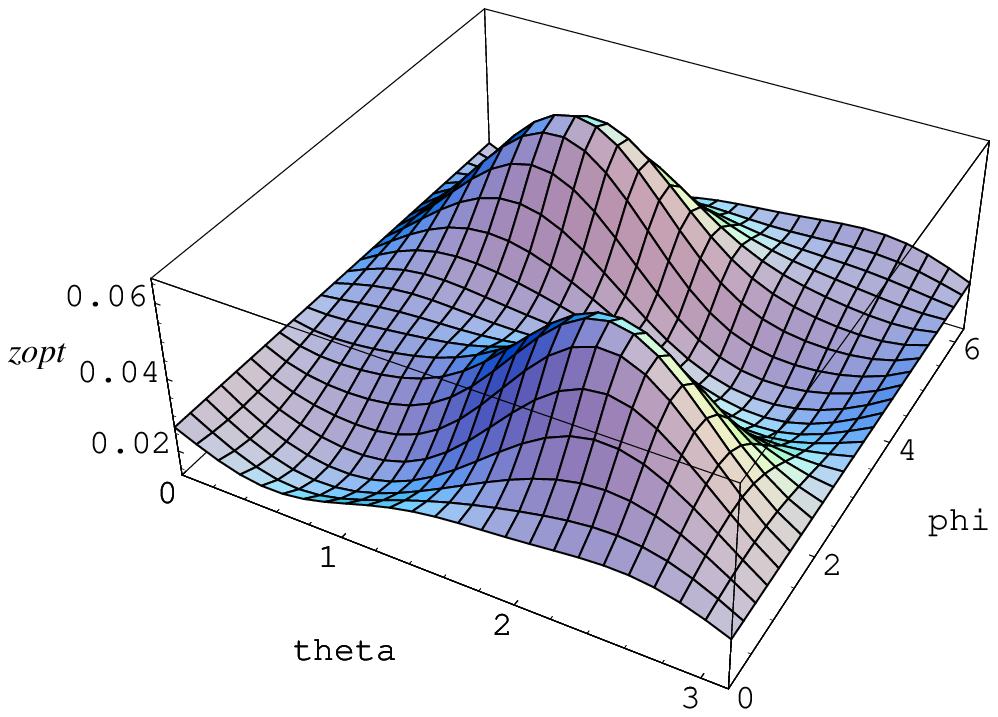,height=2.in,width=2.in}
\psfig{file=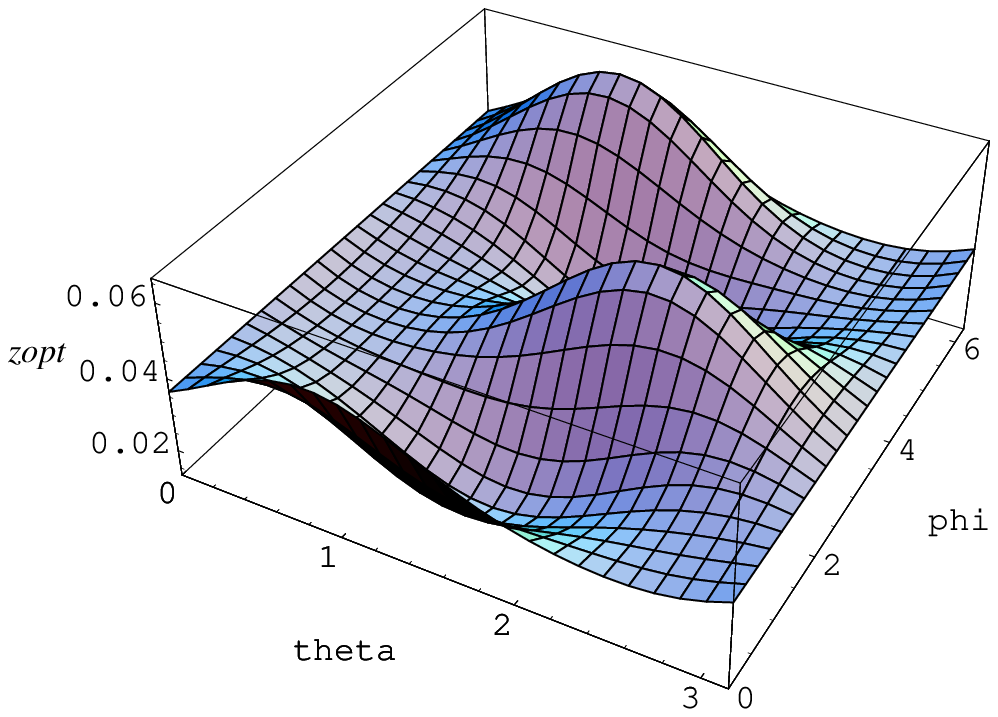,height=2.in,width=2.in}
\psfig{file=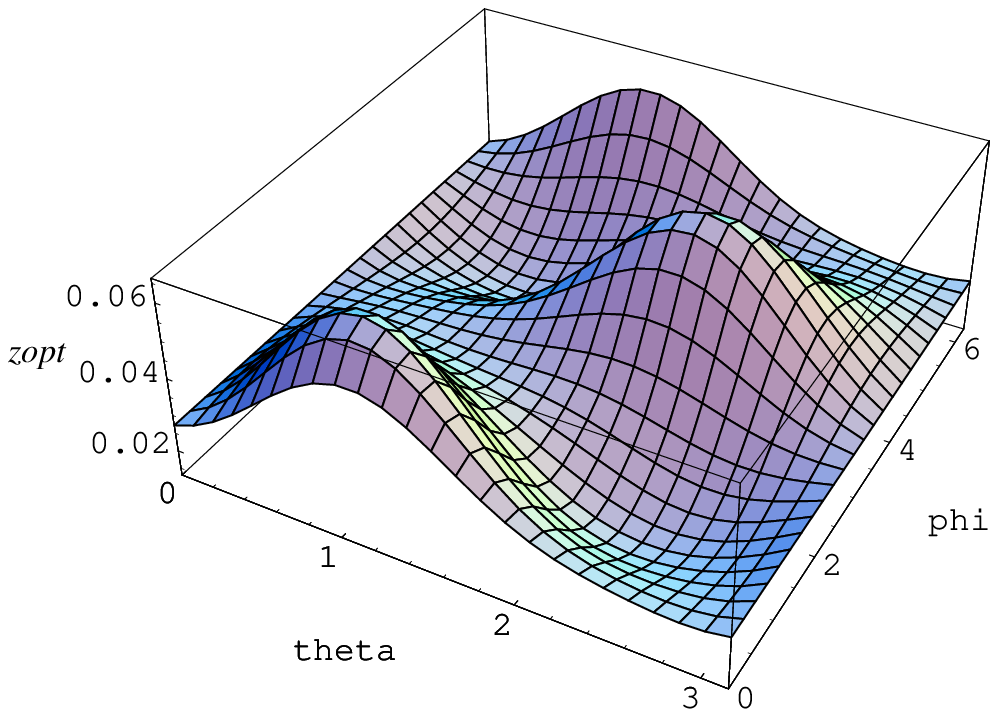,height=2.in,width=2.in}}
\caption{Network sensitivity $z_{opt}\equiv\sum_{k=1}^9 E_k^*
E^k/(g^1)^2$ evaluated at $\Omega_0 = 1$ mHz for the same orbital
positions that appear in Fig. \ref{VarySens}. It is manifest that
pseudo-detector 3 makes negligible contribution to the $z_{opt}$ at
this frequency. Note that $g^1=g^2$. At any given sky position, the
optimal statistic has better sensitivity than any $z_A$.}
\label{OptSens}
\end{figure}

In the new notation, the cross-correlation statistic becomes
\be
\label{matchedFilterk} \sum_{A=1}^{3} \langle h^A,
x^A\rangle_{(A)} = \sum_{k=1}^{9} \Re \left[e^{-{\rm i}\delta}
\hspace{3pt} E_k^* \hspace{3pt} C^{\hspace{1pt}k} \right] = \Re
\left[e^{-{\rm i}\delta}\hspace{3pt}{\bf E}\cdot {\bf C} \right] \ee
and
\be \label{signalNorm}
\sum_{A=1}^{3} \langle h^A,h^A\rangle_{(A)}
=H^2(\Omega_0)\sum_{k=1}^9 E_k^*E^k \equiv
H^2(\Omega_0) \parallel{\bf E}\parallel^2 \equiv \kappa^2 \ \ ,
\ee
where $\parallel {\bf Y}\parallel$ denotes the norm of vector ${\bf
Y}$. Therefore, $\kappa^2$ is a measure of the signal power
accessible to LISA. It is usually less than the peak power
$H^2(\Omega_0)$ owing to LISA's non-optimal orientation, ${\bf E}$,
to a given source. The relative sensitivities of the three
pseudo-detectors to different sky positions can now be studied by
plotting the analogue of $\kappa^2$ for each individual
pseudo-detector, as shown in Fig. \ref{threeSens}. This figure,
plotted for $\Omega_0=1$ mHz, verifies the fact found in Ref.
\cite{Prince:2002hp} that the third pseudo-detector has a much
smaller sensitivity than the first two. It also, shows that the {\em
peak} sensitivity of the pseudo-detector labeled as $A=1$ is the
best of the three. However, at any given location on LISA's orbit,
there are sky positions to which pseudo-detector 2 has the best
sensitivity. We also plot in Figs. \ref{VarySens} and \ref{OptSens},
the sensitivities of pseudo-detector 1 and the optimal combination
of all the pseudo-detectors, for three different locations on LISA's
orbit. These figures show that for all sky positions the
optimal-sensitivity is better than the usually best pseudo-detector,
labeled $A=1$.

The likelihood ratio now takes the following form:
\be
\ln \lambda = \kappa
\sum_{A=1}^{3}\langle \hat h^A, x^A\rangle _{(A)}-{1\over2}\kappa^2,
\ee
where
\be \hat h^A \equiv {h^A \over H(\Omega_0) \parallel {\bf E}
\parallel}
\ee
is the normalized counterpart of $h^A$, such that
$\sum_{A=1}^3 \langle \hat{h}^A, \hat{h}^A\rangle _{(A)} =1$.
The likelihood ratio can be maximized over $\kappa$ and $\delta$ to yield
\be \label{logMax}
\ln \lambda \mid_{\hat \kappa, \hat \delta} =
\frac{1}{2}\mid  \bf{Q} \cdot \bf{C} \mid^2 \ \ ,
\ee
where the hat on a parameter denotes its value at which $\ln \lambda$
stands maximized with respect to that parameter and
$\bf{Q}$ is the normalized orientation vector,
$\bf{Q} \equiv {\bf E} / \parallel {\bf E} \parallel$,
such that $\parallel {\bf Q} \parallel =1$. Also, we find
$\hat \kappa =\sum_{A=1}^3 \langle \hat{h}^A, x^A\rangle_{(A)}$ and
$\hat \delta = \arg({\bf C}\cdot {\bf Q})$.

To maximize the statistic in Eq. (\ref{logMax})
with respect to $\{\psi, \epsilon\}$,
note that ${\bf Q}$ can be expressed in terms of its components as
follows:
\be {\bf Q}\equiv Q^{+2}\hat {\bf D}_{+2} + Q^{-2}\hat {\bf
D}_{-2}, \ee where \be \hat{\bf D}_{\pm 2} \equiv {{\bf D}_\pm \over
\parallel {\bf D}\parallel} \hspace{10pt} {\rm and}
\hspace{10pt} Q^{\pm2\hspace{1pt}} \equiv{ T^{\pm 2}_2 \hspace{3pt}
\parallel {\bf D}\parallel\over\parallel {\bf E}\parallel}\,.
\ee
Above, we have used the fact that $\parallel {\bf D}_+
\parallel =
\parallel {\bf D}_-
\parallel \equiv
\parallel {\bf D} \parallel$.
The statistic in Eq. (\ref{logMax}) depends on
$\{\psi,\epsilon\}$ solely through ${\bf Q}$. Therefore, it stands
maximized with respect to those parameters when ${\bf Q}$ gets
aligned with ${\bf C}$. The fact that this alignment is physically
realizable was shown in Ref. \cite{Bose:1999pj}. The
maximized statistic is
\be \label{max3} \ln \lambda |_{\hat
\kappa, \hat \delta, \hat \psi, \hat \epsilon} = \frac{1}{2}
\parallel \bf{C}_\mathcal{H}\parallel^2 \ \ ,
\ee where $\bf{C}_\mathcal{H}$ is the projection of $\bf{C}$ on a
2-dimensional complex space, $\mathcal{H}$, spanned by $\{\hat {\bf
D}_{+2}, \hat {\bf D}_{-2}\}$. Since it is always possible to choose
a pair of real basis vectors to define this two-dimensional space,
we take these vectors to be 
\be 
\hat{\bf o}^\pm \equiv
\left(\hat{\bf d}_1 \pm \hat{\bf d}_2\right) /
\parallel \hat{\bf d}_1 \pm \hat{\bf d}_2 \parallel \ \ ,
\ee where ${\bf d}_1 = \Re({\bf D}_{+2})$ and ${\bf d}_2 = \Im({\bf
D}_{+2})$. Thus, we may define the network search statistic as
\be\label{netStat} \Lambda \equiv \parallel {\bf
C}_\mathcal{H}\parallel^2 = |C^+|^2 + |C^-|^2 =
(c_0^+)^2+(c_{\pi/2}^+)^2+(c_0^-)^2+(c_{\pi/2}^-)^2 \ \ , \ee where
$C^\pm = \hat{\bf o}^\pm\cdot {\bf C} \equiv c_0^\pm +
ic_{\pi/2}^\pm$. The maximizing values of the two parameters are
$\hat \psi = \arg(\varpi)/4$ and $ \hat \epsilon = \cos^{-1}
\left[(1-\sqrt{|\varpi|})/(1+\sqrt{|\varpi|})\right]$, where $\varpi
\equiv C^{+2}/ C^{-2}$.

The above statistic is still a function of the sky position through the
parameters $\{ w_1, w_2 \}$. Since it is not possible to maximize the
statistic over these parameters analytically, one must resort to a numerical
maximization scheme as described in the following section. By comparing the
values of the statistic for each pixel in the sky with a threshold value,
$\Lambda_0$, a decision on the presence or absence of
a signal in the data can be made. The threshold itself is determined by
the false-alarm probability that one can afford.
Note that in the absence of a signal,
$\Lambda$ is a random variable that has a $\chi^2$ probability distribution,
\be
p_0(\Lambda) = \frac{\Lambda}{4}\exp\left(-\Lambda/2\right) \ \ ,
\ee
with 4 degrees of freedom \cite{Pai:2000zt}. This is because each of the
$c_0^\pm$ and $c_{\pi/2}^\pm$ is a Gaussian random variable with a zero
mean and a unit variance. The false-alarm probability is
\be
Q_0 = \int_{\Lambda_0}^\infty p_0(\Lambda)d(\Lambda)=\left(1+\frac{\Lambda_0}{2}\right)
\exp\left(-\Lambda_0/2\right)\,.
\ee
In the presence of a signal, the probability distribution of $\Lambda$ is
non-central $\chi^2$,
\be
p_1(\Lambda) = \frac{\sqrt{\Lambda}}{2\kappa}\exp\left(-\frac{\Lambda+\kappa^2}{2}\right)
I_1\left(\kappa\sqrt{\Lambda}\right) \ \ ,
\ee
with the non-centrality parameter as $\kappa^2$,
which is a measure of the signal power \cite{Pai:2000zt}. Above, $I_1$
is the modified Bessel function.

\section{\label{sec:templates}Template Spacing and Computational Costs}

The $\Lambda$ statistic in Eq. (\ref{max3}) must be maximized over
the remaining intrinsic parameters, namely,  $\mbox{\boldmath
$\vartheta$} = \{\Omega_0, \dot{\Omega}_0, w_1, w_2\}$. 
As a first calculation, we will set $\dot{\Omega}_0=0$ and will focus 
on the number of templates required to scan the space of 
$\{\Omega_0, w_1, w_2\}$ for this case. 
As noted above, the ensuing template spacings will still be relevant to a 
large number of compact binaries that involve white-dwarfs. The corrections
arising to these spacings for non-zero $\dot{\Omega}_0$ will be studied
elsewhere. The maximization can then 
be achieved numerically using a discrete template bank over this
three-dimensional parameter space. The drop in the
value of the statistic and, therefore, the
signal-to-noise ratio (SNR) that one can
afford determines how coarsely one can space the templates. In
practice, there are limits posed by the available computational resources
on how fine the spacing can be. The loss in SNR is related to
the template spacing through the ambiguity function.\cite{Hels}

The ambiguity function corresponding to the $\Lambda$ statistic is
derived from it by replacing the data $x^A$ there by a signal $h^A
(\mbox{\boldmath $\vartheta$}')$. We distinguish the signal
parameter values from those of the template by denoting the former
with a prime. The parameter values of a template used in a search
may not be the same as those of a signal hiding in the data. Let the
parameter mismatch be $\Delta\mbox{\boldmath $\vartheta$} \equiv
(\mbox{\boldmath $\vartheta$}' -\mbox{\boldmath $\vartheta$})$. Then
the ambiguity function is a real quantity expressed as 
\be
\label{simplesquarestatistic} m(\mbox{\boldmath $\vartheta$}, \Delta
\mbox{\boldmath $\vartheta$}) \equiv p_k^{~l}Q^{'k\hspace
{2pt}}Q^{'*}_{l}\Theta_{(k)(l)}, 
\ee 
where $Q^{'k\hspace {2pt}}$
depends only on the signal parameters and 
\bea \Theta_{(k)(l)}
&=&\langle S^{' \hspace{2pt}k}, S^{k} \rangle^*_{(k)} \langle S^{'
\hspace{2pt}l}, S^{l} \rangle_{(l)} \no\\&=& e^{{\rm i} \Psi_k}
e^{{\rm i} \Psi_l} {\rm sinc}(\Omega_{k}^{'}T-\Omega_{k}T){\rm
sinc}(\Omega_{l}^{'}T-\Omega_{l}T), 
\eea 
and $p_k^{~l} \equiv
o^{+l}o_k^+ + o^{-l}o_k^-$ is an amplitude factor. It is important to
note that as $\Delta \mbox{\boldmath $\vartheta$} \to 0$, one has
$ m(\mbox{\boldmath $\vartheta$}, \Delta \mbox{\boldmath $\vartheta$})
\to1$, which is the maximum value it can attain.
The correlation phase, $\Psi_k$, is defined as
\be \label{correlationphase} \Psi_k = L(\Omega_{k}^{'}
\tau^{'}_{k} -\Omega_{k} \tau_{k})=\Omega_{k} {\hat{r}_{k1}\over
2\sqrt{3}}\Delta w_{1}+ \Omega_{k} {\hat{r}_{k2}\over
2\sqrt{3}}\Delta w_{2} - \tau_{k}\Delta \Omega_k. \ee
The drop in the value of $m(\mbox{\boldmath $\vartheta$},
\Delta \mbox{\boldmath $\vartheta$})$ caused by non-zero, but small
$\Delta \vartheta^\mu$, can be ascertained
by Taylor expanding it about the maximum at
$\Delta \mbox{\boldmath $\vartheta$} = 0$. \cite{Owen:1995tm,Bose:1999pj}
The first order term is zero since by
definition the statistic has a maximum when the template
parameters match the signal parameters. Thus,
\be
1-m(\mbox{\boldmath $\vartheta$},
\Delta \mbox{\boldmath $\vartheta$}) \simeq \gamma_{\alpha\beta}
\Delta \vartheta^\alpha\Delta \vartheta^\beta \ \ ,
\ee
$\gamma_{\alpha\beta}$ is determined from the second order term in
that expansion:
\be
\label{metric} \gamma_{\alpha \beta} ={ -{1\over
2}\left({\partial^2 m(\mbox{\boldmath $\vartheta$},
\Delta \mbox{\boldmath $\vartheta$})\over \partial \Delta \vartheta^\alpha
\partial \Delta \vartheta^\beta}
\right)} \Bigg |_{\Delta \mbox{\boldmath $\vartheta$}= 0}. \ee
It defines the metric on the 3-dimensional parameter space.

The computational cost for the search can be reduced by taking advantage
of the Fast Fourier Transform algorithms \cite{numRecipes} and computing the
cross-correlation components, ${\bf C}$, in the Fourier domain. This
defines the strategy for searching for the source frequency, $\Omega_0$.
To search for the remaining parameters, $\{w_1,w_2\}$, one must design a
bank of ``templates'' with values of these sky positions spaced such that
the loss in SNR is never more than the desired fraction, say, $\mu$. To
find the metric, $g_{ij}$, on the two-dimensional space $\mathcal{P}$,
spanned by $\{w_1,w_2\}$,
we project $\gamma_{\alpha\beta}$ orthogonal to $\Omega_0$,
\be \label{reducedmetric} g_{ij}=
\gamma_{i\hspace{1pt} j}-{\gamma_{0\hspace{1pt}
i}\gamma_{0\hspace{1pt} j} \over \gamma_{0\hspace{1pt} 0}} \ \ ,\ee
where $i$ and $j$ span only the $\{w_1,w_2\}$ space, and the
index $0$ denotes the $\Omega_0$ axis. The volume of a $\mathcal{P}$
is then given by
\be\label{volume}
\mathcal{V}=\int_{\mathcal{P}}\sqrt{{\rm det}\|g_{a
\hspace{1pt} b}\|}d^P \vartheta \ \ , \ee
where $P=2$ is the dimensionality of the space.
The number density of templates, $\rho_P(\mu)$,
is determined as a function of $\mu$ to be \cite{Bose:1999pj}:
\be \label{numberdensity}
\rho_P(\mu)=\left({1\over 2} \sqrt{P\over \mu} \right)^P \,.\ee
Setting the fractional SNR loss $\mu= 3$\%
yields a template density of
$\rho_2(0.03)=16.6$. Therefore, the total number of templates is
just the overall parameter volume times the template density, i.e.,
$N_{\rm templates}=\mathcal{V}\cross \rho_P(\mu)$.

\begin{figure}[!hbt]
\centerline{\psfig{file=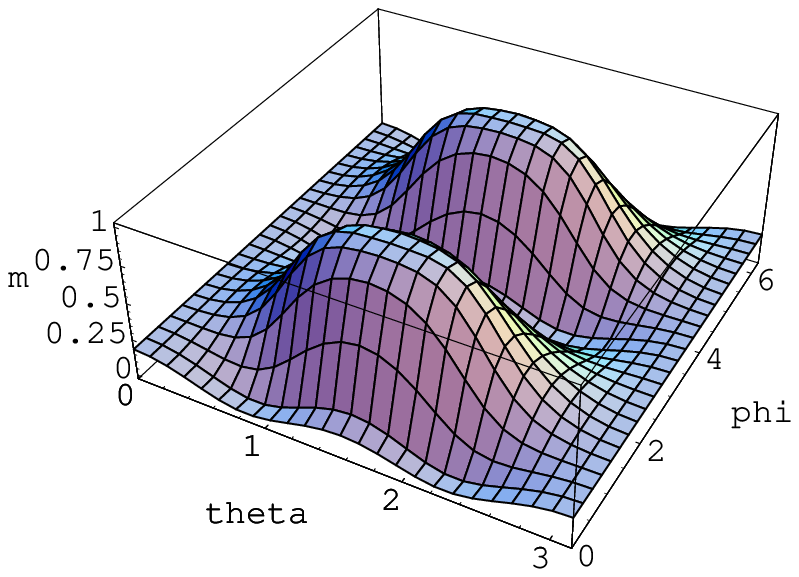,height=2.in,width=2.in}
\psfig{file=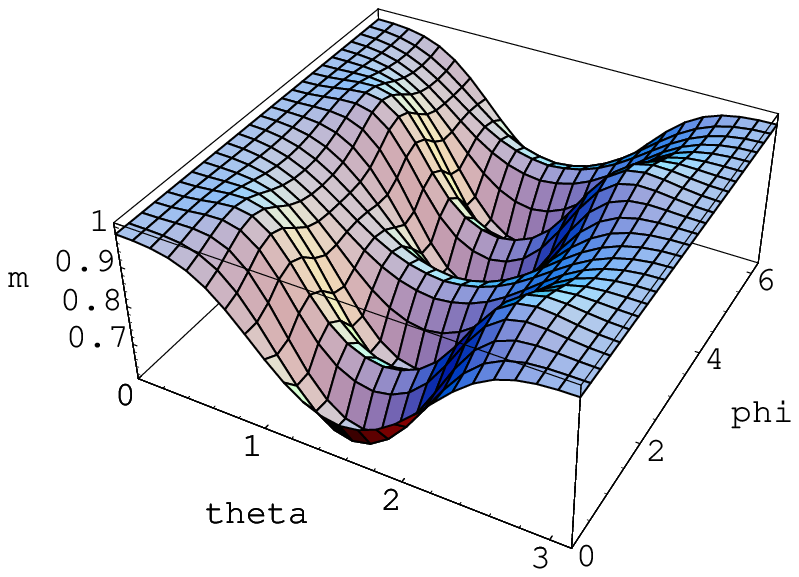,height=2.in,width=2.in}
\psfig{file=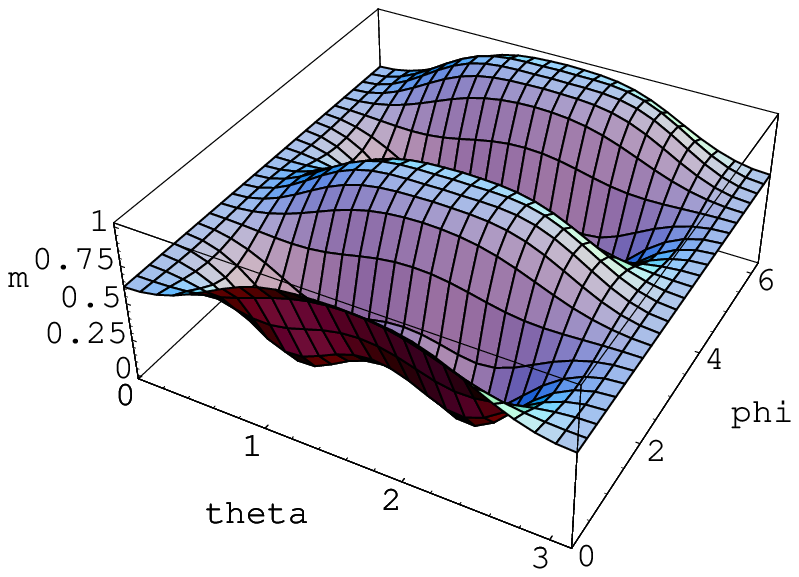,height=2.in,width=2.in}}
\caption{The ambiguity function $m$ plotted as a function of the
template parameters, $\{\theta,\phi\}$, for three different source
sky positions (in radians), namely, $\{\theta',\phi'\} =
\{\pi/2,4\pi/3\}$, $\{\pi/6,\pi/3\}$, and $\{\pi/2,\pi\}$.}\label{fig:ambFun}
\end{figure}

The parameter volume, obtained via the metric computation, turns out
to be about 5 for $\Omega_0=1$ mHz. Considering that the template
density per unit volume is only 16.6 implies that the number of
sky-position templates required for a search with 3\% loss of SNR
is about 80. The smallness of this number is not surprising,
given how slowly the ambiguity function varies as a function of
$\Delta \theta$ and $\Delta \phi$, as shown in Fig. \ref{fig:ambFun}.

As is manifest from
Eq. (\ref{simplesquarestatistic}), in principle,
this variation can arise from either the time delays in $\Theta_{kl}$ or
the weights $p_k^{~l}$.
However, for wavelengths much larger than the LISA arm length,
the ability to discern between different sky positions through the
time delays in $\Theta_{kl}$ is negligible. The main contribution to
$m(\mbox{\boldmath $\vartheta$}, \Delta \mbox{\boldmath $\vartheta$})$,
therefore, arises from the $p_k^{~l}$. For detailed studies of the
angular resolution achievable by LISA, we refer the reader to
Refs.  \cite{Cutler:98,Takahashi:02}.

To get a handle on the computational costs associated with a search
of this nature, one must determine the overall number of sampling
points in a single data train. For our analysis, we have chosen a
sampling rate of 1 Hz. Therefore, the number of sampling points is
just the length of the data train, $T$. The number of floating point
operations associated with an FFT with $T$ sampling points is: \be
\label{FFT} N_{\rm fl-opts}=T \hspace{2pt} {\rm log}_2 T. \ee
Considering that this must be done for each template, the resulting
expression for the total number of floating point operations for an
arbitrary template bank is just $N_{\rm templates} \cross N_{\rm
ft-opts}$. However, the real quantity of interest is the the number
of operations per second, \be \label{flops} C^{\rm flops} \equiv
{N_{\rm templates} \cross N_{\rm ft-opts}\over T}=N_{\rm
templates}{\rm log}_2 T \,. \ee Therefore, the computational costs
of implementing a real-time search with about 80 templates on a
year's worth of data is trivial since the number of flops scales
logarithmically with the integration time.

\section{\label{sec:summary}Discussion}

In this paper, we developed an optimal method for detecting 
slightly chirping compact-binary inspiral signals in the LISA data. 
We also studied the geometrical properties and sensitivities of the three
noise-independent pseudo-detectors or data combinations of LISA.
Following the earlier work on TDI data combinations, it was found 
\cite{Cornish:2003tz,Shaddock:2003dj,Tinto:2003vj} 
that the rotational motion of LISA and the time-variation of
LISA's arm-lengths would prevent the noise contribution of the 
laser-frequency fluctuations from being mitigated to the level of the 
secondary noises. In order to tackle this problem, second generation
pseudo-detectors were introduced as simple differences of their first 
generation counterparts, appropriately time-shifted:
\be\label{pseudo2}
{x}^{\bar{A}}(t) \equiv x^A(t)-x^A(t-L_1-L_2-L_3)\simeq x^A(t)-x^A(t-3 L) \,.
\ee
The analysis presented above can be easily extended for detecting inspiral
signals in the second generation pseudo-detectors, ${x}^{\bar{A}}(t)$, by 
implementing the next two steps: 
First, by following the derivation in Sec. \ref{sec:pseudo} 
it can be verified 
that the noise PSDs of the new detectors are given in terms of the old ones by:
\be\label{noisePSD2}
P^{(\bar{1},\bar{2},\bar{3})}(f) = 4\sin^2(3\pi fL) P^{(1,2,3)}(f)\,.
\ee
Second, since a GW contribution to ${x}^{\bar{A}}(t)$ will also get 
differenced as in Eq. (\ref{pseudo2}), the template that must be 
matched against ${x}^{\bar{A}}(t)$ 
should itself be modified accordingly:
\be
{S}^{\bar{A}}_i(t) \equiv S^A_i(t)-S^A_i(t-L_1-L_2-L_3)
\simeq S^A_i(t) - S^A_i(t-3L) \,.
\ee
None of the formal analysis presented in this paper is affected by this change.
For example, the analytic maximization of the likelihood ratio over 
four source parameters (namely, the signal strength, the initial phase, 
the polarization angle and the orbital-inclination angle), 
and the concomitant computational gain achieved in the process, still hold 
so long as the matched-filter output in Eq. (\ref{signal}) is redefined to be
\be
C^{\bar{A}}_i =
\langle {S}_i^{\bar{A}}, {x}^{\bar{A}}\rangle_{(\bar{A})} 
\,. \ee
Also, the sensitivity plots in Figs. \ref{threeSens}, \ref{VarySens},
and \ref{OptSens} for the first generation
detector ${A}$ are the same as those of its second generation counterpart,
i.e., detector $\bar{A}$. This is because the (geometric) sensitivity, $z_A$, 
depends on the orientation of a detector relative to the source
and is independent of the noise PSDs.  
Similarly, the formal expressions for the ambiguity function in
the last section above remain unchanged. But, since this function depends
on the noise PSDs, its numerical value is affected by the change in 
Eq. (\ref{noisePSD2}). Nevertheless, we have found that this change 
has negligible effect on the number of templates deduced in 
Sec. \ref{sec:templates}.

A useful by-product of our analysis is that it yields the 
maximum-likelihood estimates of the initial phase, the polarization angle,
and the angle of inclination (in addition to, of course, the signal 
amplitude). To complete the parameter estimation problem, however, one needs
to derive the errors associated with them as well as in the parameters that
will be searched for numerically, viz., $\{\Omega_0,\dot{\Omega}_0,
\theta,\phi\}$. 
That problem will be addressed elsewhere.

\ack

We would like to thank Rajesh Nayak for helpful correspondence. We
are grateful to Massimo Tinto for pointing out to us that in the
long-wavelength approximation, the contribution from the GW strain
to third pseudo-detector discussed here should become negligible. The
pseudo-detector sensitivities plotted in Fig. \ref{threeSens}
verify this claim. We would also
like to thank Curt Cutler for critically reading the manuscript and 
apprising us of Ref. \cite{Jaranowski:1998qm}, which is a closely related 
work on the data analysis of gravitational-wave signals from spinning 
neutron stars for Earth-based laser interferometric detectors.
This work was funded in part NASA Grant NAG5-12837.


\begin{thebibliography}{9}
\bibitem{ceilingMod}
Although widely used by programmers, the operations of Ceiling and Modulus
may not be familiar to most physicists and astronomers. The Ceiling of
$x$ is denoted as $\lceil x \rceil$ and
is equal to the smaller integer not less than $x$.
E.g., for $k=2$, the Ceiling($k/3$) equals 1; and for $k=6$, it is 2.
On the other hand, $x$ modulus $y$ is expressed as $x$\%$y$ and is
equal to the remainder when $x$ is divided by $y$. E.g., 2\%3 = 2 and
6\%3=0.
\end{thebibliography}
\end{document}